\documentclass[prd,aps,twocolumn,preprintnumbers,amsmath,amssymb,nofootinbib,superscriptaddress,notitlepage]{revtex4-1}
\newcommand{\ii}{\mathrm{i}}
\newcommand{\ee}{\mathrm{e}}
\newcommand{\dd}{\mathcal{d}}

\usepackage{amsmath, amssymb, amsthm, amsfonts, booktabs, extarrows, moreenum, color, mathtools, url, bm, enumitem, graphicx, mathrsfs, subcaption, physics, tensor, slashed}
\usepackage[normalem]{ulem}

\usepackage{hyperref}
\hypersetup{hidelinks,
	colorlinks=true,
	allcolors=blue,
	pdfstartview=Fit,
	breaklinks=true}

\usepackage[mathscr]{euscript}

\begin{document}

\title{Resonance sum rules: an application to the square well potential
}

 \author{Zi-Xi Ou-Yang}%
\email[Email: ]{ouyangzixi23@mails.ucas.ac.cn}
\affiliation{School of Physical Sciences, University of Chinese Academy of Sciences (UCAS), Beijing 100049, China}

\author{Philipp Gubler}%
\email[Email: ]{philipp.gubler1@gmail.com}
\affiliation{Advanced Science Research Center, Japan Atomic Energy Agency, Tokai 319-1195, Japan}

\author{Makoto Oka}%
\email[Email: ]{makoto.oka@riken.jp}
\affiliation{Nishina Center for Accelerator-Based Science, RIKEN, Wako 351-0198, Japan}
\affiliation{Advanced Science Research Center, Japan Atomic Energy Agency, Tokai 319-1195, Japan}

\author{Guang-Juan Wang}%
\email[Email: ]{wgj@post.kek.jp}
\affiliation{KEK Theory Center, Institute of Particle and Nuclear Studies (IPNS), High Energy Accelerator Research Organization (KEK), Tsukuba 305-0801, Japan}

\author{Jia-Jun Wu}%
\email[Email: ]{wujiajun@ucas.ac.cn}
\affiliation{School of Physical Sciences, University of Chinese Academy of Sciences (UCAS), Beijing 100049, China}

\date{\today}%
\begin{abstract}
We propose an extension of the Quantum Chromodynamics (QCD) sum rules, termed the Resonance sum rules (RSR), to access resonance poles in the complex energy plane. 
By strategically introducing a contour in the complex plane and conformal mapping, the method intends to reach resonance poles on the second Riemann sheet. 
To validate this approach, we apply RSR to the square-well potential model, for which the pole locations are known. 
The analysis demonstrates a successful extraction of the pole positions and residues for both the $S$-wave and $P$-wave resonances. 
The results are in good agreement with the analytic solutions, with discrepancies within $5\%$ for the pole positions and $20\%$ for the residues.
This framework provides a basis for future applications to realistic hadronic resonances, promising new insights into spectral properties of QCD.
\end{abstract}
\maketitle

\section{Introduction }
\label{sec:intro}

The QCD sum rules method provides a powerful theoretical framework to study hadronic properties directly from the fundamental principles of Quantum Chromo-Dynamics (QCD), the theory of the strong interaction.
Originally developed by Shifman, Vainshtein, and Zakharov~\cite{Shifman:1978bx, Shifman:1978by}, this approach connects hadronic properties with quark and gluon dynamics through operator product expansion (OPE), dispersion relations, and quark-hadron duality. 
By analyzing the correlation function, 
\begin{align}
\Pi(q^2)&=\ii\int \dd^{4}x \ee^{\ii q\cdot x}\langle \mathsf{T}[j(x)j^{\dagger}(0)] \rangle, 
\label{eq:correlator}
\end{align}
constructed from a local current operator, $j(x)$, of quarks and/or gluons, one can extract hadronic parameters such as masses, decay constants, and form factors from the underlying QCD dynamics.
The non-perturbative contributions encoded in vacuum condensates, such as the quark and gluon condensates, serve as essential inputs to the formalism. 
Despite being semi-phenomenological, the QCD sum rules approach has provided substantial insights into hadron spectroscopy and structure, especially for systems where lattice QCD or other non-perturbative techniques become computationally intensive or less effective. 
For comprehensive reviews, see Refs. \cite{Reinders:1984sr,Leinweber:1995fn,Colangelo:2000dp,Qiao:2014vva,Gubler:2018ctz,Wan:2021vny,Wang:2025sic}. 

A central challenge lies in the treatment of resonances, which constitute a dominant part of the hadron spectrum.  
In lattice QCD, the discrete energy levels of a finite volume just provide the scattering amplitudes at these real energy points.
Then, using theoretical models or effective field theory, the scattering amplitude is analytically continued into the complex energy plane to find the resonance pole position. 
This process not only requires significant computational resources to obtain the energy levels, but also introduces substantial theoretical uncertainties due to the models used for determining the pole position.
By contrast, QCD sum rules method exploits the analyticity of correlation functions to connect the OPE in the deep Euclidean region to the physical kinematic region in the complex energy plane. 
This suggests that sum rules could provide a complementary approach to exploring the analytic structure of resonant states.

However, the conventional framework still has critical limitations. 
The physical spectral function constrained by the sum rules method is defined only on the positive real axis of the square of four-momentum ($s=q^2$), while the resonance poles reside in the complex plane.
Moreover, the spectral function is widely modeled via the ``pole + continuum" ansatz~\cite{Shifman:1978bx, Shifman:1978by} 
\begin{align}
\rho(s)&\equiv \frac{1}{\pi}\mathrm{Im} \Pi(s) \nonumber\\
&= |\lambda|^2 \delta(s - m^2) + \frac{1}{\pi} \theta(s - s_{\mathrm{th}}) \mathrm{Im} \Pi^{\mathrm{OPE}}(s), 
\label{eq:pole_cont}
\end{align}
where $m$ is the mass of the hadron, which is assumed to have a vanishing decay width.
$\lambda$ denotes the coupling of current $j(x)$ to the same state. 
Finally, $\Pi^{\mathrm{OPE}}(s)$ is the OPE expression of the original correlator in Eq.~(\ref{eq:correlator}), which can be expected to be accurate at high energy and hence is used to approximate the continuum above the threshold $s_{\mathrm{th}}$. 
This assumption, called ``quark-hadron duality'', is justified in QCD thanks to asymptotic freedom. 
The ansatz in Eq.~(\ref{eq:pole_cont}), is well justified for a narrow ground state below the continuum threshold, while it becomes inadequate for resonances with finite decay widths.

Efforts have been made to improve these shortcomings. 
A finite-width spectral peak has been introduced into the spectral function with a variable width parameter to be determined by the sum rules~\cite{Leupold:1997dg, Morita:2007pt, Erkol:2008gp,Chen:2007xr}, or the integral of the sum rules is directly inverted to access the spectral function without using Eq.~(\ref{eq:pole_cont}) ~\cite{Gubler:2010cf, Li:2020ejs}.
These studies have revealed that the standard sum rule framework cannot fully constrain all three resonance parameters (mass, coupling, and width) simultaneously, but instead leads to certain correlations among them. 
For example, a Maximum Entropy analysis of the $\rho$ meson sum rule~\cite{Gubler:2010cf} did reveal the existence and location of a peak in the spectral function, but could not unambiguously determine its width. 
Thus, the conventional sum rules framework remains limited in its ability to access the full pole structure of resonances.

To overcome these limitations, we explore a generalization of QCD sum rules in the complex energy plane. 
Inspired by the complex scaling method used in many-body systems~\cite{Myo:2014ypa,Myo:2020rni,Moiseyev:1998gjp}, we explore a modification of the dispersion relation, 
\begin{align}
\Pi(q^2) =  \frac{1}{\pi} \oint_{\mathbb C} \dd s \frac{\mathrm{Im}\Pi(s + \ii\varepsilon)}{s - q^2},
\label{eq:disp}
\end{align}
by deforming the integration contour $\mathbb C$ into the complex plane.
This allows for an analytic continuation of the correlator onto the second Riemann sheet, thus enabling the extraction of pole positions associated with resonant states.

Before applying this idea to QCD, we test it in 
a simple and analytically solvable system {\it i.e.,} the quantum mechanics of the square well potential.
Depending on its depth, this potential can exhibit bound, virtual, and resonant states. 
Each corresponds to a pole of the scattering amplitude in the complex energy plane and leaves its (more or less pronounced) characteristic traces on the real axis.  
Our aim is to examine whether complex-plane sum rules can extract the resonance poles in the low energy region from the asymptotic behavior of the scattering amplitude at large momentum. 

The paper is organized as follows. 
In Section~\ref{sec:SWP}, we introduce the square well potential and its spectrum.
Section~\ref{sec:boundstate} outlines the sum rule formulation for extracting the bound state, while in Section~\ref{sec:resoance}, we present the novel method to determine the pole position of resonance states and their residues.
In the next section, we show how to calculate the resonance parameters using the above method for the $S$- and $P$-wave cases.
Conclusions are given in Section \ref{sec:concl}, and technical derivations are provided in the Appendices. 

\section{Square Well Potential }
\label{sec:SWP}

We consider a 3-dimensional square-well potential, 
\begin{equation}V(R)=\begin{cases}
	-V_0 &R<\mathcal{R}\\
	0 &R>\mathcal{R}\\
\end{cases},
\end{equation}
where $R$ is the radial coordinate, and $V_0$ and $\mathcal{R}$ are the depth and the range of the potential, respectively. 
The radial wave function with the angular momentum $l$ reads
\begin{equation}
    \psi_l(R)=\begin{cases}
	c_l^V j_l(k_1R),&R<\mathcal{R}\\
	c_l^+ h_l^{(+)}(kR)+c_l^- h_l^{(-)}(kR),&R>\mathcal{R}\\
\end{cases},
\end{equation}
where
\begin{equation} k_1\equiv\sqrt{k^2+k_0^2},\,\, k\equiv\sqrt{2mE},\,\,k_0\equiv\sqrt{2mV_0},
\end{equation}
and $h_l^{(\pm)}=j_l\pm\ii n_l$ are the spherical Hankel functions, with $j_l$ and $n_l$ being the spherical Bessel and Neumann functions, respectively.

Given the continuity of the wave function at $R=\mathcal{R}$, the scattering amplitude matrix $T_l$ is defined by
\begin{equation}
    T_l(E)\equiv\frac{S_l(z)-1}{2\ii z},
\end{equation}
where the matrix $S_l(z)$ is
\begin{equation}
    S_l(z)\equiv\frac{c_l^+}{c_l^-}=\frac{zh_l^{(-)\prime}(z)j_l(z_1)-z_1h_l^{(-)}(z)j_l'(z_1)}{z_1h_l^{(+)}(z)j_l'(z_1)-zh_l^{(+)\prime}(z)j_l(z_1)},
\end{equation}
with dimensionless variables $z=k\mathcal{R},\ z_1=k_1\mathcal{R}=\sqrt{z^2+z_0^2}$, where $z_0=k_0\mathcal{R}$.
For simplicity, we set $\mathcal{R}=1$ and $m=1$. 
Then, one has $E=\frac{1}{2}z^2$, $V_0=\frac{1}{2}z_0^2$. 
The bound state exists if $z_0\geqslant \pi/2$ for $l=0$, and $z_0\geqslant \pi$ for $l=1$.

\begin{figure*}
    \begin{minipage}{0.48\textwidth}
        \centering
        \includegraphics[width=0.8\columnwidth]{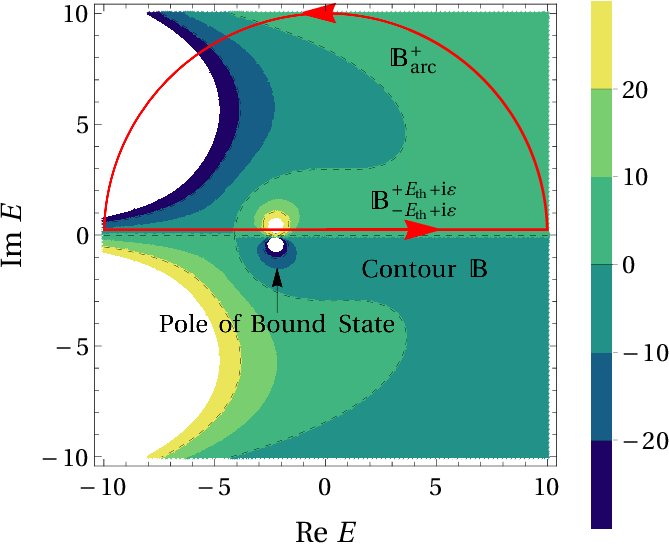}
        \subcaption{RI}
    \end{minipage}
    \begin{minipage}{0.48\textwidth}
        \centering
        \includegraphics[width=0.8\columnwidth]{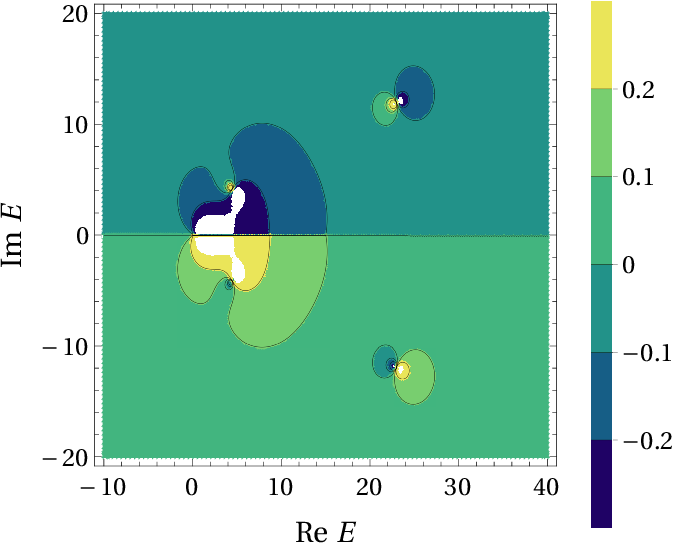}
        \subcaption{RII}
    \end{minipage}
    \caption{Contour plot of imaginary part of $S$-wave scattering amplitude on the first (RI, left) and second (RII, right)  Riemann sheets of the complex energy plane for $z_0=\pi$. On the RI sheet, the red closed path $\mathbb{B}$, denotes the integration contour that excludes the bound state pole on the negative real axis.}
    \label{Fig:ScatteringAmplitude}
\end{figure*}

For $l=0$, the scattering amplitude is explicitly given by
\begin{equation}\label{sazero}
    T_0(z)=\frac{(z\sin{z_1}\cos{z}-z_1\cos{z_1}\sin{z})}{z( z_1\cos{z_1}-\ii z\sin{z_1})}\ee^{-\ii z},
\end{equation}
which is analytic in the complex $z$-plane except for the poles corresponding to the physical states.
The square root in the momentum introduces two Riemann sheets: the first (physical) (RI) sheet contains bound states, while resonant and virtual states exist on the second Riemann sheet (RII).
Fig.~\ref{Fig:ScatteringAmplitude} shows the imaginary parts of the scattering amplitude with $z_0=\pi$ on the RI and RII sheets.

\section{Extraction of Bound state Poles} \label{sec:boundstate}

In this section, we present a method for extracting the real poles, i.e., the energy of the bound states, generated from the square-well potential. 
As their decay widths vanish, the analysis is similar to that of conventional QCD sum rules studies.
This hence serves as a test of the usefulness of the square-well potential model as an analogue of real QCD sum rules.

\subsection{Large energy expansion of $T$-matrix}

For large energies $E$, the scattering amplitude $T_l(E)$ can be approximated by a series $T_l^{\mathrm{OPE}(n)}(E)$ with truncation order $n$, which represents a power series in $V_0/|E|$. 
The expansion is valid in the limit $V_0/|E| \to 0$ and therefore becomes increasingly accurate for $|E|\gg V_0$, corresponding to the weak-coupling or high-energy limit of the system.

\begin{figure*}
    \begin{minipage}{0.48\textwidth}
        \centering
        \includegraphics[width=0.9\columnwidth]{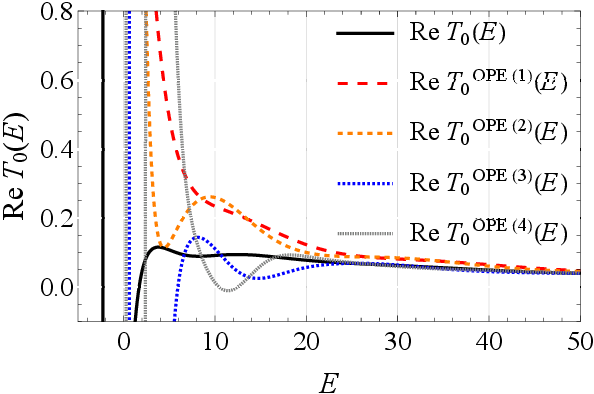}
        \subcaption{Real Part}
    \end{minipage}
    \begin{minipage}{0.48\textwidth}
        \centering
        \includegraphics[width=0.9\columnwidth]{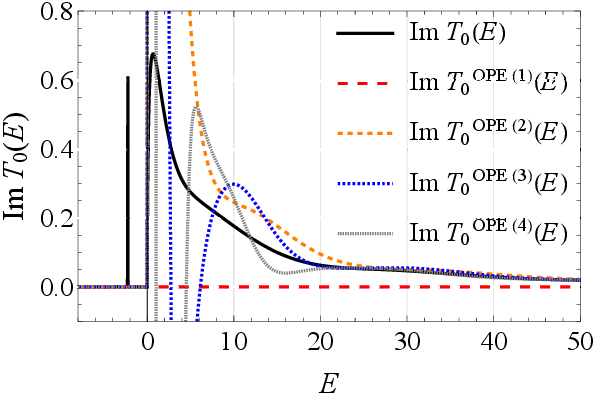}
        \subcaption{Imaginary Part}
    \end{minipage}
    \caption{Comparison of the exact $S$-wave scattering amplitude $T_0(E)$ with the truncated expansion $T_0^{\mathrm{OPE}(n)}(E)$ on the first Riemann Sheet for $z_0=\pi$ ($V_0=\pi^2/2$). The black solid line refers to $T_0(E)$ and $T_0^{\mathrm{OPE}(n)}(E)$ with $n=1,2,3,4$ are shown by the red, orange, blue, and gray dashed lines, respectively.}
    \label{Fig:OPEShow}
\end{figure*}

Although there are no explicit operators here, the expansion can be considered as the analogue of the OPE in QCD sum rules.\footnote{The OPE is an expansion by the mass dimension of the local operators, and thus the Wilson coefficients scale inversely with this dimension. 
Such feature can be mimicked by the present expansion.}
As an explicit example, for $l=0$, the expansion reads 
\begin{align}
   &T_0(z)=(2z-\sin{2z})\frac{z_0^2}{4z^3}+(4\sin{2z}+\ii-\ii \ee^{4\ii z}\nonumber\\
    &\quad-4z-8\ee^{2\ii z}z+8\ii z^2)\frac{z_0^4}{32z^5}+\cdots.
\end{align}
We define $T_l^{\mathrm{OPE}(n)}(E)$ by truncating this series at the order $(z_0^2)^{n}$ or $V_0^{n}$.
In Fig.~\ref{Fig:OPEShow}, we compare the exact scattering amplitude $T_0(E)$ with its truncated expansions $T_0^{\mathrm{OPE}(n)}(E)$ for $z_0=\pi$ ($V_0=\pi^2/2$). 
The expansion reproduces the exact amplitude for sufficiently large $|E|$ along the real axis and the convergence systematically improves with increasing truncation order $n$. 

\subsection{Pole extraction}

To extract the bound state poles of the $T$-matrix, we consider a contour integral on the first Riemann sheet of the complex energy plane RI, as shown in the left panel of Fig.~\ref{Fig:ScatteringAmplitude}.
The closed contour $\mathbb{B}$ consists of a straight line parallel to the real axis (shifted by an infinitesimal $+\ii\varepsilon$), and an arc $\mathbb{B}_\mathrm{arc}^+$ closing the contour in the upper half-plane:
\begin{equation}
    \oint_\mathbb{B}=\int_{-E_{\mathrm{th}}+\ii\varepsilon}^{E_{\mathrm{th}}+\ii\varepsilon}+\int_{\mathbb{B}^+_{\mathrm{arc}:E_{\mathrm{th}}}},
\end{equation}
where $\mathbb{B}^{+/-}_{\mathrm{arc}:E_{\mathrm{th}}}$ denote anti-clockwise/clockwise integrals along the arc of radius $E_{\mathrm{th}}$.

The contour $\mathbb{B}$ excludes the bound state poles on the negative real axis of $E$ in the first Riemann sheet.
As a result, for any analytic scattering amplitude $T(E)$, the integral
\begin{equation}\label{Equ:ContourIntegral}
    \oint_\mathbb{B}T(E)W(E,M)\dd E=0,
\end{equation}
vanishes by Cauchy's theorem. 
Here, $W(E,M)$ is an analytic weight function in the complex energy $E$ plane and becomes a real number along the real energy $E$ axis, with $M$ being a parameter called ``Borel mass''.

Defining the spectral function $\rho_l(E)=\frac{1}{\pi}\mathrm{Im}T_l(E+\ii\varepsilon)$, Eq.~\eqref{Equ:ContourIntegral} implies~\cite{Chen:2013zia,Qiao:2014vva}
\begin{align}
\mathcal{L}_l(M,E_{\mathrm{th}},n)&\equiv\mathrm{Im}\int_{\mathbb{B}_{\mathrm{arc}:E_{\mathrm{th}}}^-}T_l^{\mathrm{OPE}(n)}(E)W(E,M)\dd E\nonumber\\
&\approx\int_{-E_{\mathrm{th}}}^{E_{\mathrm{th}}}
\pi\rho_l(E)W(E,M)\dd E.\label{eq:LL}
\end{align}
Notably, although $T_l^{\mathrm{OPE}(n)}(E)$ is valid and convergent only for sufficiently large $|E|$ (arc section of the integral), the analyticity ensures that the contour integral in Eq.~\eqref{Equ:ContourIntegral} 
relates high-energy expansion to low-energy pole information.

On the phenomenological side, assuming that a single bound state dominates, the imaginary part of the $T$-matrix can be parameterized as 
\begin{equation}
   \mathrm{Im}T_l^\mathrm{ph}(E+\ii\varepsilon)=\mathrm{Im}\left[\frac{-f_l}{E-E_l+\ii\varepsilon}+\mathrm{background}\right],\label{eq:tbd}
\end{equation}
where $E_l$ and $f_l$ are the energy and the coupling constant of the bound state, respectively. 
The “background” here represents all other contributions that do not originate from this bound state, including the continuum and the high energy states above the threshold $E_{\mathrm{th}}$.
Consequently, the spectral function is then
\begin{equation}
    \rho^\mathrm{ph}_l(E)= f_l\delta(E-E_l)+\rho_l^{\mathrm{bg}}(E)\theta(E-E_{\mathrm{th}}),\label{eq:tbdrho}
\end{equation}
where $\rho_l^{\mathrm{bg}}(E)$ describes the contribution of ``background".

Assuming $\rho^\mathrm{ph}_l=\rho_l$ and using the contour relation in Eq.~(\ref{eq:LL}), one obtains the integration of $\rho^\mathrm{ph}_l$ from $T^{\mathrm{OPE}}_l$. 
To enhance the ground-state contribution and suppress the ``background" contribution, we choose a weight function of the form $W(E,M) = \ee^{-E/M}$. 
The extracted energy and coupling constant are\cite{Kojo:2006bh,Kojo_2008,Chen_2011,Chen:2013zia,Qiao:2013raa,Kim_2020}
\begin{align}
E_l(M,E_{\mathrm{th}},n)&=M^2\frac{\mathrm{d}}{\mathrm{d} M}\ln{\mathcal{L}_l(M,E_{\mathrm{th}},n)},\label{eq:El}\\
f_l(M,E_{\mathrm{th}},n)&=\frac{1}{\pi}\ee^{E_l/M}\mathcal{L}_l(M,E_{\mathrm{th}},n).\label{eq:fl}
\end{align}
From Eqs.~(\ref{eq:El}-\ref{eq:fl}), the extracted bound state energy and coupling constant depend on three parameters: the Borel mass $M$, the continuum threshold $E_{\mathrm{th}}$, and the truncation order $n$. 

The Borel window and the threshold $E_{\mathrm{th}}$ must be appropriately constrained. 
To this end, we employ the standard criteria used in QCD sum rules: large pole contribution (PC) and good OPE convergence (CVG).
This is because, on one hand, if $M$ is too large, the ``background" contribution becomes significant. 
To suppress it, we require the PC~\cite{Kojo_2008,Chen:2013zia,Chen_2011,Kim_2020,Yang:2024okq}
\begin{equation}
\mathrm{PC}=\frac{\mathcal{L}_l(M,E_{\mathrm{th}},n)}{\mathcal{L}_l(M,\infty,n)},
\end{equation}
of the lowest bound state to be dominant. 
On the other hand, if $M$ is too small, CVG then cannot be guaranteed. 
To address the problem above, the ratio below should exceed a lower bound, i.e.~\cite{Kojo_2008,Chen:2016otp,Kim_2020,Yang:2024okq}
\begin{equation}
    \mathrm{CVG}=\left|1-\frac{\mathcal{L}_l(M,E_{\mathrm{th}},n-1)}
    {\mathcal{L}_l(M,E_{\mathrm{th}},n)}\right|,
\end{equation}
where $n$ stands for the truncation order of OPE.

Although the numerical thresholds for PC and CVG are not uniquely determined, they must be stringent enough to ensure pole dominance and OPE convergence. 
The $E_{\mathrm{th}}$ is then fixed by minimizing the $M$-dependence of the extracted energy through~\cite{Kim_2020,Yang:2024okq} 
\begin{equation}\label{Euq:XiSquaredBound}
\chi^2(E_{\mathrm{th}})=
\int_{M_\mathrm{min}}^{M_\mathrm{max}}\mathrm{d}M
\frac{[E_l(M,E_{\mathrm{th}},n)-\bar{E}_l]^2}{M_\mathrm{max}-M_\mathrm{min}},
\end{equation}
where
\begin{equation}
\bar{E}_l=\int_{M_\mathrm{min}}^{M_\mathrm{max}}\mathrm{d}M\frac{E_l(M,E_{\mathrm{th}},n)}{M_\mathrm{max}-M_\mathrm{min}}.
\end{equation}

As an illustrative example, we consider a trivial case with $l=0$, $z_0=\pi$ and $\rho_0^{\mathrm{OPE}(n=10)}$. 
Due to the large coupling $f_0$, one has $\mathrm{PC}>95\%$ even for $M>100$, leading to a wide Borel mass range.  
Nevertheless, the upper bound $M_{\mathrm{max}}=10$ is imposed to ensure $\mathrm{PC}\geqslant 98\%$. 
For the CVG criterion, imposing $\left.\mathrm{CVG}\right|_{n=10}<3\times10^{-3}$, leads to the lower limit of the Borel window $M_{\mathrm{min}} = 4.5$. 
By minimizing $\chi^2$, we find the optimal threshold $E_{\mathrm{th}}\approx5.66$. 

\begin{figure*}
    \begin{minipage}{0.48\textwidth}
        \centering
        \includegraphics[width=0.9\columnwidth]{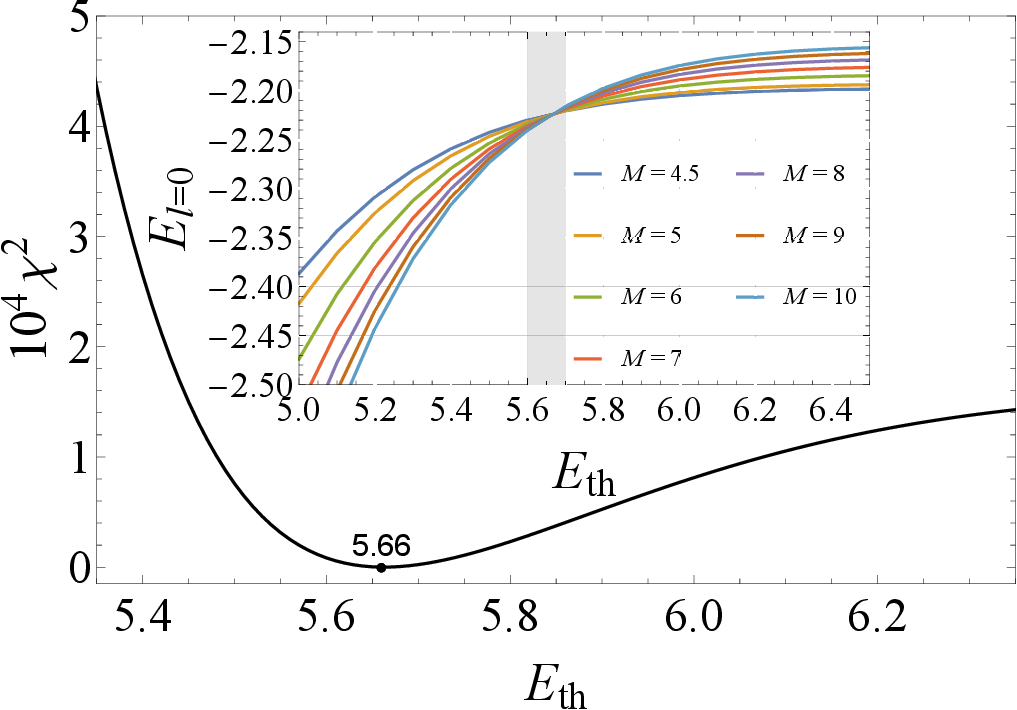}
        \subcaption{Determination of the continuum threshold $E_{\text{th}}$ by  minimizing  $\chi^2$ in Eq. \eqref{Euq:XiSquaredBound} (main panel). The inset shows an alternative determination by locating the intersection point of Borel curves $E_{l=0}$ versus $E_{\text{th}}$ for different values of $M$. Both methods give consistent results. } \label{Fig:expEM_1_th}
    \end{minipage}
    \begin{minipage}{0.48\textwidth}
        \centering
        \includegraphics[width=1.0\columnwidth,height=0.202\textheight]{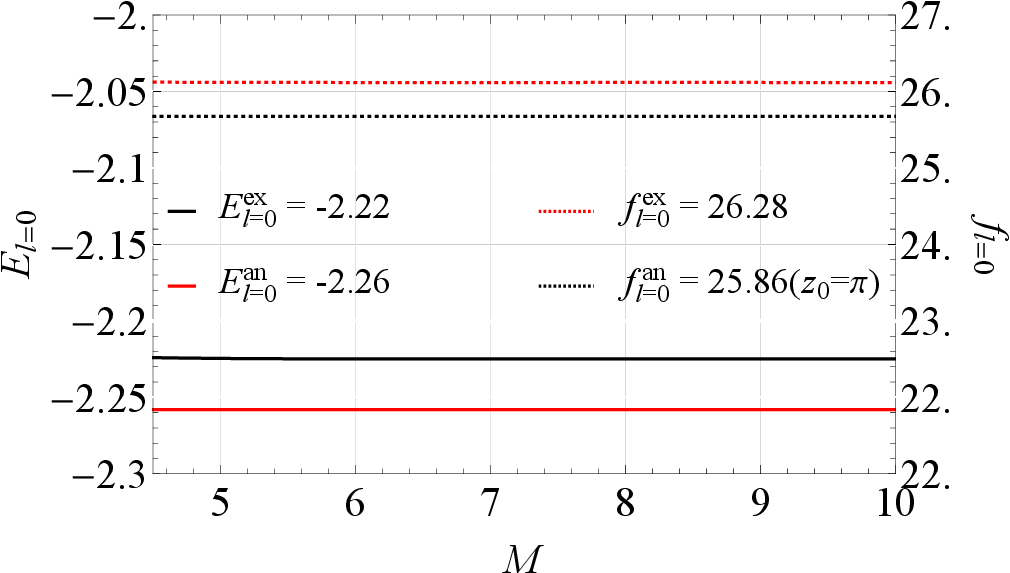}
        \subcaption{Bound state energy and residue as functions of the Borel mass $M$, obtained using the optimal threshold $E_{\mathrm{th}}\approx5.66$. 
        The extracted values (red) are reasonably close to the analytic ones (blue). } \label{Fig:expEM_1_energy_residue}
    \end{minipage}
    \caption{Extraction of the $S$-wave bound state pole for $z_0=\pi$ ($V_0=\pi^2/2$) from OPE up to truncation order $n=10$ using a method analogous to conventional QCD sum rules.}
    \label{Fig:S-wave}
\end{figure*}

For comparison, another common and equivalent approach is to determine $E_{\mathrm{th}}$ from the intersection point of Borel curves $E_{l=0}$ versus $E_{\mathrm{th}}$ for different values of $M$~\cite{Chen_2011,Yang:2024okq}. 
As shown in Fig.~\ref{Fig:expEM_1_th}, the minimum of $\chi^2$ coincides with the intersection point of the Borel curves. 
This confirms the consistency of the determination of $E_{\mathrm{th}}$ and demonstrates that the extracted energy is insensitive to $M$.

Finally, the extracted bound state energy $E_{l=0}^\mathrm{ex}=-2.22$ and the coupling constant $f_0^\mathrm{ex}=26.28$ are in agreement with the analytic results $E_{l=0}^\mathrm{an}=-2.26,\ f_0^\mathrm{an}=25.86$, respectively, as illustrated in Fig.~\ref{Fig:expEM_1_energy_residue}.
Here, for simplicity, the superscripts ``$\mathrm{ex}$"  and ``$\mathrm{an}$" are for `extracted' and `analytic', respectively.
The relative deviations, $(\mathrm{analytic}-\mathrm{extracted})/\mathrm{analytic}$ , are approximately $1.5\%$ for the energy and $1.6\%$ for the coupling constant.

\section{Resonance Sum Rules to Extract the Complex Pole Position}
 \label{sec:resoance}
 
\begin{figure*}
    \begin{minipage}{0.48\textwidth}
        \centering
        \includegraphics[width=0.9\columnwidth]{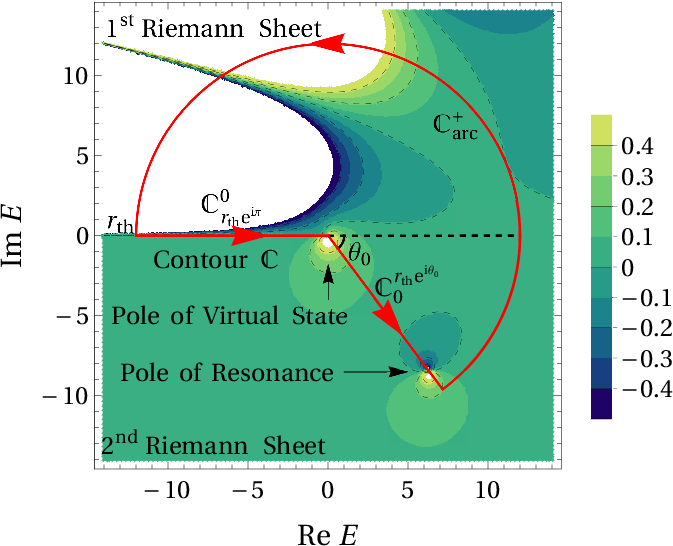} 
        \subcaption{Contour  $\mathbb{C}$ in the complex-energy plane employed in the new resonance sum rule for the $z_0=0.4\pi$ case. }  \label{Fig:ContourIntegral2}
    \end{minipage}
    \begin{minipage}{0.48\textwidth}
        \centering
        \includegraphics[width=0.9\columnwidth]{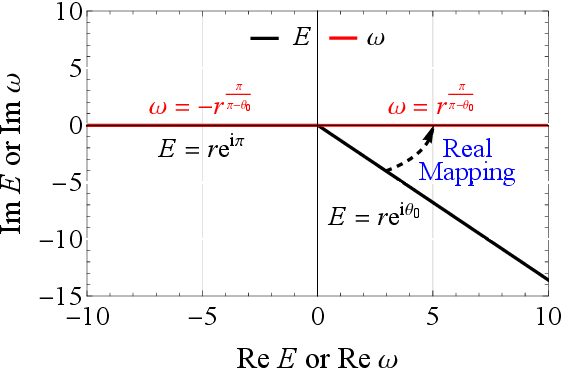}
        \subcaption{Example of the conformal mapping $\omega(E,\theta_0)$ in Eq. \eqref{eq:Etoomega}, which maps the contour segments   $E=r\ee^{\ii\pi}$ and $E=r\ee^{\ii\theta_0}$ to the real $\omega$ axis for $\theta_0=-0.936$.} \label{Fig:EToOmega}
    \end{minipage}
    \caption{Illustration of the contour and conformal mapping used in the resonance sum rule.}
    \label{Fig:rsm}
\end{figure*}

In this section, we extend the conventional sum rule framework to extract the resonance pole. 
It is necessary to determine both the real and imaginary parts of the complex pole position on the second Riemann sheet. 
This introduces several challenges beyond those encountered in the bound state case: 
(i) the integration contour must be deformed to access the second Riemann sheet, 
(ii) the linear segments of the contour must lie close to the direction specified by the phase angle of the pole, so that the integration path passes through the vicinity of the resonance, and
(iii) both the real and imaginary parts of the $T$-matrix must be utilized. 
To overcome these challenges, we introduce a new contour and mapping procedure, as shown in Fig.~\ref{Fig:rsm}, which allows us to determine the phase angle and modulus of the pole simultaneously.

\subsection{$T$-matrix near a complex pole}

On the phenomenological (ph) side, the $T$-matrix containing a resonance pole can be written as
\begin{equation}
    T^\mathrm{ph}(E)=\frac{F}{E-E_P}+\mathrm{background}, \label{eq:Tph}
\end{equation} 
with the residue $F=f_x+\ii f_y$ and the pole position $E_P=r_P\ee^{\ii\theta_P}$ ($\theta_P<0$) \footnote{The second Riemann sheet corresponds to the angular domain $\theta\in(-2\pi,0)$.} and complex argument $E=r\ee^{\ii\theta}$.
At the vicinity of the pole $\theta\to \theta_P+0^\pm$, one has
\begin{equation}\label{eq:TT}
\begin{split}
    \left.T^\mathrm{ph}(E)\right|_{\theta\to\theta_P+0^\mp}=&F\ee^{-\ii\theta}\left[\frac{1}{r-r_p}\pm\ii\pi\delta(r-r_P)\right]\\
    &+\mathrm{background}. 
\end{split}
\end{equation}

Unlike the bound state case described in Eq.~\eqref{eq:tbdrho}, the imaginary part alone cannot isolate the pole contribution $\delta(r-r_P)$, since both the pole position and the residue are complex for a resonance.
Thus, the conventional sum rules method is no longer applicable.

\subsection{New method to extract the complex pole}

To resolve the above complications, we propose a novel method to extract the complex pole based on Eq.~(\ref{eq:TT}).
First, we deform the contour connecting the asymptotic region with the vicinity of the pole, as shown in Fig.~\ref{Fig:ContourIntegral2}. 
The closed contour includes three parts: $\mathbb{C}^0_{r_{\mathrm{th}}\ee^{\ii\pi}}$ on RI along its negative axis, $\mathbb{C}_0^{r_{\mathrm{th}}\ee^{\ii\theta_0}}$ on RII and $\mathbb{C}^+_{\mathrm{arc}}$ from RI to RII crossing the positive axis of RI at which RI and RII connect, and then yields,
\begin{align}
&\oint_{\mathbb{C}}T(E)W(E,M)\mathrm{d}E
\nonumber \\
\quad\quad
    =&\left\{\begin{array}{cc}
        0 & \,\,\,\,\theta_0 >\theta_P \\
        2\pi\ii FW(r_P\ee^{\ii\theta_P},M) & \,\,\,\,\theta_0<\theta_P
    \end{array}\right.,\label{eq:TCinte}
\end{align}
where $\mathbb{C}^{+/-}_{\mathrm{arc}:E_{\mathrm{th}}}$ denote anti-clockwise/clockwise integrals along an arc of radius $E_{\mathrm{th}}$. 
The paths $\mathbb{C}^0_{r_{\mathrm{th}}\ee^{\ii\pi}}$ and $\mathbb{C}_0^{r_{\mathrm{th}}\ee^{\ii\theta_0}}$ are mapped onto the real axis by the transformation,  
\begin{align}\label{eq:Etoomega}
    \omega(E,\theta_0)=(E\ee^{-\ii\theta_0})^\frac{\pi}{\pi-\theta_0},
\end{align}
where $\theta_0$ is the newly defined scanning angle. 
Under this mapping, $ \omega(E,\theta_0)$ takes $-r^{\pi/(\pi-\theta_0)}$ on the contour $\mathbb{C}_{r_{\mathrm{th}}\ee^{\ii\pi}}^0$ for $E=r \ee^{\ii\pi}$ and $r^{\pi/(\pi-\theta_0)}$ on $\mathbb{C}^{r_{\mathrm{th}}\ee^{\ii\theta_0}}_0$ for $E=r \ee^{\ii\theta_0}$, as illustrated in Fig.~\ref{Fig:EToOmega}. 

The weight function is chosen to have a Gaussian form,
\begin{align}\label{eq:Etowf}
    W(E,M;E_0,\theta_0)=\exp \left(-\frac{[\omega(E,\theta_0)-\omega(E_0,\theta_0)]^2}{M^2}\right)
\end{align}
where $E_0=r_0\ee^{\ii\theta_0}$ is a new parameter that changes the peak of the Gaussian weight function to effectively enhance the contribution from the vicinity of the target pole position. (We hereafter suppress the $(E_0,\theta_0)$ from the argument of $W$ for simplicity.)

When $\theta=\theta_0\approx\theta_P$ and $r\approx r_P$, one has $T(E)\approx T^{\mathrm{ph}}(E)$ along $\mathbb{C}_0^{r_{\mathrm{th}}\ee^{\ii\theta_0}}$. 
Moreover, an appropriate Gaussian weight suppresses the ``background" contribution of the integral along the path $\mathbb{C}^0_{r_{\mathrm{th}}\ee^{\ii\pi}}$, which can thus be neglected. 
Hence, we define
\begin{align}
&A+\ii B
\equiv \left(\int^0_{r_{\mathrm{th}}\ee^{\ii\pi}}+\int_0^{r_{\mathrm{th}}\ee^{\ii\theta_0}}\right)T(E)W(E,M)\dd E\nonumber\\ 
&\xlongequal[{\theta_0\to\theta_P}]{\mathrm{ph}}
\int_{0}^{r_{\mathrm{th}}}
(f_x+\ii f_y)\Bigg[\frac{1}{r-r_P}\pm \ii\pi\delta(r-r_P)\Bigg]\nonumber\\
&\quad\quad\quad\times W(r\ee^{\ii\theta_0},M)\mathrm{d}r, \label{Equ:A+iB}\\
& C+\ii D
\equiv\left(\int^0_{r_{\mathrm{th}}\ee^{\ii\pi}}+\int_0^{r_{\mathrm{th}}\ee^{\ii\theta_0}}\right)T(E)W(E,M)\omega(E,\theta_0)\dd E\nonumber\\
& \xlongequal[{\theta_0\to\theta_P}]{\mathrm{ph}}
\int_{0}^{r_{\mathrm{th}}}
(f_x+\ii f_y)\Bigg[\frac{1}{r-r_P}\pm \ii\pi\delta(r-r_P)\Bigg]\nonumber\\
&\quad\quad\quad\times W(r\ee^{\ii\theta_0},M)\omega(r\ee^{\ii\theta_0},\theta_0)\mathrm{d}r,\label{Equ:C+iD}
\end{align}
where $A, C$ and $B, D$ denote the real and imaginary parts of integrals, respectively. 
Here and in the following, the variables with ``$\mathrm{ph}$" indicate that $T^\mathrm{ph}(E)$ is applied in the case that angle of $E=r\ee^{\ii\theta}$ along $\mathbb{C}_0^{r_\mathrm{th}\ee^{\ii\theta_0}}$ is in the vicinity of $\theta_0\approx\theta_P$, and the ``background" term is assumed to be neglected.  
The ``$\pm$" signs stand for the cases $\theta_0<\theta_P<0$ and $\theta_P<\theta_0<0$, respectively.

To eliminate the contribution of the $1/(r-r_P)$ term, we introduce a new parameter $\alpha$ such that $f_x+\alpha f_y = 0$. 
This leads to simple relations, 
\begin{align}
&(A+\alpha B)^\mathrm{ph}_{\theta_0\to\theta_P}=\pm\pi(\alpha f_x-f_y)W(r_P\ee^{\ii\theta_0},M), \label{Equ:APlusAlphaB}\\
&(C+\alpha D)^\mathrm{ph}_{\theta_0\to\theta_P}=\pm\pi(\alpha f_x-f_y)W(r_P\ee^{\ii\theta_0},M)\nonumber\\
&\qquad \times\omega(r_P\ee^{\ii\theta_0},\theta_0).
\label{Equ:CPlusAlphaD}
\end{align}
Here, the ``$\pm$" signs are inherited from Eqs.~(\ref{Equ:A+iB}-\ref{Equ:C+iD}). 
This implies that $A+\alpha B$ and $C+\alpha D$ exhibit discontinuity when $\theta_0$ passes through $\theta_P$. 
The derivation of Eqs.(\ref{Equ:APlusAlphaB}-\ref{Equ:CPlusAlphaD})  can be found in Appendix~\ref{Appendix:Phenomenological Derivation}.

Next, we define the arc integrals as
\begin{align}\label{eq:abprime}
&    A'+\ii B'\equiv\int_{\mathbb{C}^- _\mathrm{arc}}T(E)W(E,M)\mathrm{d} E,
\end{align}
\begin{align}\label{eq:cdprime}
&    C'+\ii D'\equiv\int_{\mathbb{C}^-_\mathrm{arc}}T(E)W(E,M)\omega(E,\theta_0)\mathrm{d} E.
\end{align}
With the residue relation in Eq.~\eqref{eq:TCinte}, $(A',B')$ and $(C',D')$ can be related to $ (A,B)$ and $(C,D)$, respectively. 
In particular, one finds 
\begin{equation}\begin{split}
&(A'+\alpha B')^\mathrm{ph}_{\theta_0\to\theta_P}=\pi(f_y-\alpha f_x)W(r_P\ee^{\ii\theta_P},M),
\end{split}\end{equation}
\begin{equation}\begin{split}
&( C'+\alpha D')^\mathrm{ph}_{\theta_0\to\theta_P}=\pi(f_y-\alpha f_x)W(r_P\ee^{\ii\theta_P},M)r_P^\frac{\pi}{\pi-\theta_P}.
\end{split}\end{equation}
Notably, the above relations hold regardless of whether the resonance pole lies inside or outside the closed contour $\mathbb{C}$. 
The detailed derivation of the above equations is given in Appendix~\ref{Appendix:Phenomenological Derivation}.

When the scanning angle $\theta_0$ approaches the pole angle $\theta_P$ ($\theta_0\to\theta_P$), the pole position can be extracted directly as
\begin{equation}\label{eq:EPget}
E_P^\mathrm{ex}=r_P^\mathrm{ex}\ee^{\ii\theta_0},
\end{equation}
\begin{equation}
r_P^\mathrm{ex}=\left(\frac{C'+\alpha D'}{A'+\alpha B'}\right)^\frac{\pi-\theta_0}{\pi},
\end{equation}
where $E^\mathrm{ex}$ and $r^\mathrm{ex}$, denote the extracted complex energy and its modulus, 
and coincide exactly with the true pole position $E_P^\mathrm{ex}|_{\theta_0\to\theta_P}=E_P^\mathrm{an}=r_P^\mathrm{an}\ee^{\ii\theta_P}$ if all ``background" contributions vanish.

Finally, the residue $F=f_x+\ii f_y$ can be extracted as
\begin{equation}\label{eq:FExtraction}
    F^\mathrm{ex}=\frac{(A'+\alpha B')^\mathrm{ph}_{\theta_0\to\theta_P}}{\pi W(E_P^\mathrm{ex},M)}\frac{-\alpha+\ii}{\alpha^2+1},
\end{equation}
by solving the equations
\begin{equation}
    \left\{\begin{array}{c}
         f_x+\alpha f_y=0,  \\
         (A'+\alpha B')^\mathrm{ph}_{\theta_0\to\theta_P}=\pi(f_y-\alpha f_x)W(E_P^\mathrm{ex},M).
    \end{array}\right.
\end{equation}

\subsection{Determination of parameters}

\begin{figure*}
    \centering
    \begin{minipage}{0.49\textwidth}
        \centering
        \includegraphics[width=0.8\columnwidth]{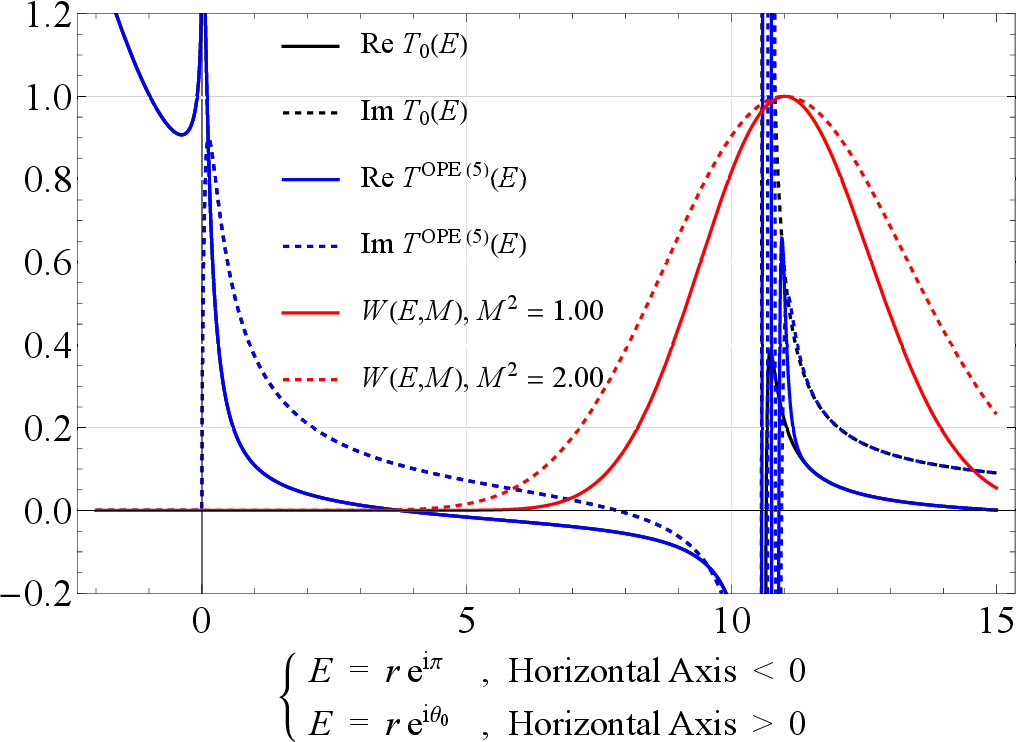}
    \subcaption{S-wave}
    \end{minipage}
    \begin{minipage}{0.49\textwidth}
        \centering
        \includegraphics[width=0.8\columnwidth]{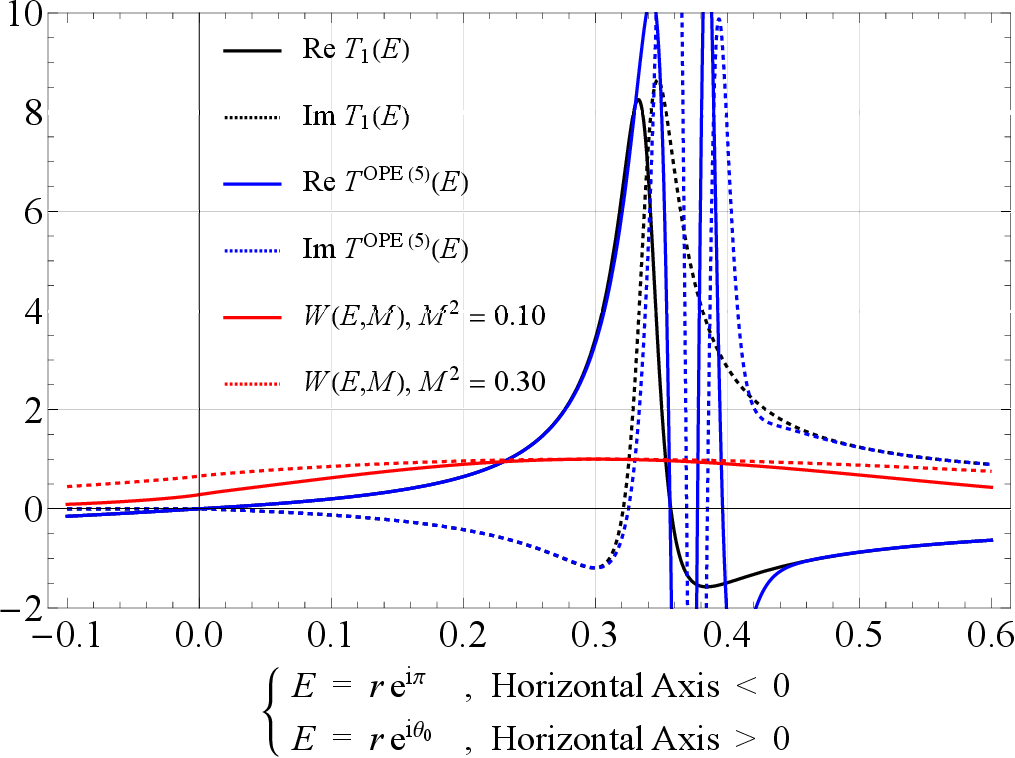}
    \subcaption{P-wave}
    \end{minipage}
    \caption{Scattering amplitudes, their truncated expansions, and applied weight functions within the new resonance sum rule for the S-wave (upper) and P-wave (lower) channels.} \label{Fig:ScatteringAmplitudeAndWeightFunction}
\end{figure*}

\begin{figure*}
    \centering
    \begin{minipage}{0.49\textwidth}
        \centering
        \includegraphics[width=0.9\textwidth]{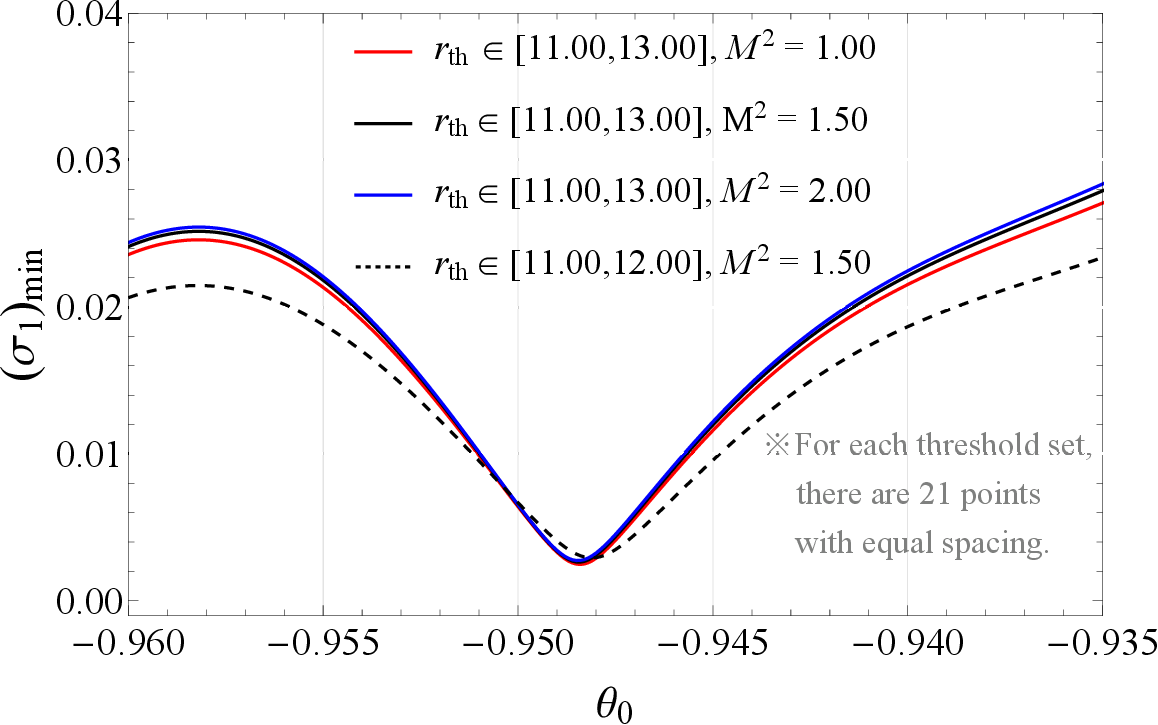}
        \subcaption{S-wave}
    \end{minipage}
    \begin{minipage}{0.49\textwidth}
        \centering
        \includegraphics[width=0.9\textwidth]{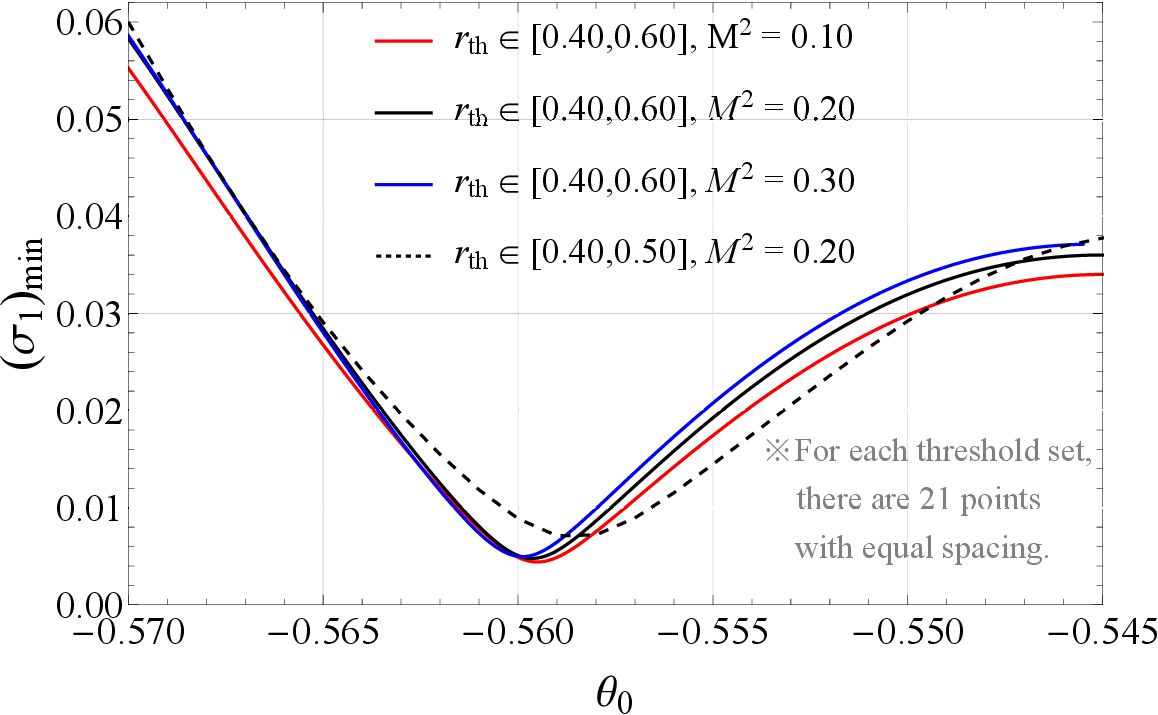}
        \subcaption{P-wave}
    \end{minipage}
    \caption{Minimum value of $\sigma_1$ by varying $\alpha$, shown as a function of $\theta_0$.}
    \label{Fig:ThetaAndSigma}
\end{figure*}

\begin{figure*}
    \centering
    \begin{minipage}{0.49\textwidth}
        \centering
        \includegraphics[width=0.9\textwidth]{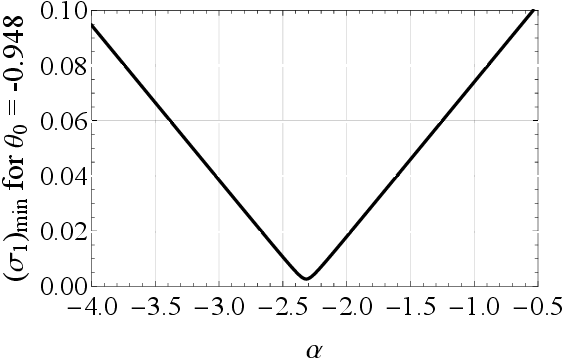}
        \subcaption{S-wave}
    \end{minipage}
    \begin{minipage}{0.49\textwidth}
        \centering
        \includegraphics[width=0.9\textwidth]{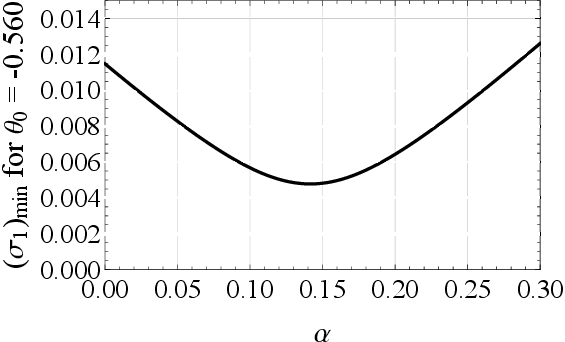}
        \subcaption{P-wave}
    \end{minipage}
    \caption{The value of $\sigma_1$ as a function of $\alpha$ with the fixed $\theta_0$.}
    \label{Fig:alphaandsigma}
\end{figure*}

\begin{figure*}
    \centering
    \begin{minipage}{0.49\textwidth}
        \centering
        \includegraphics[width=0.95\columnwidth]{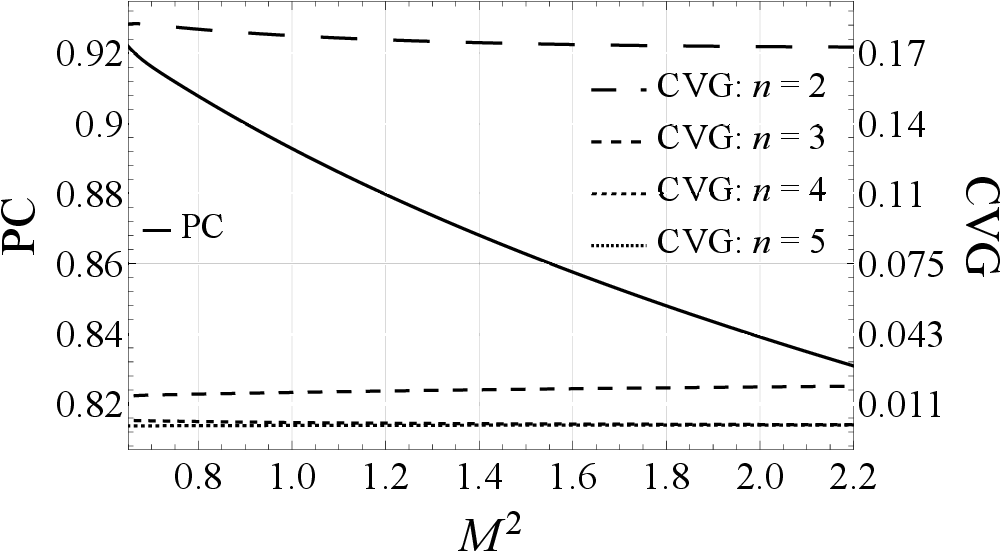}
        \subcaption{S-wave}
    \end{minipage}
    \begin{minipage}{0.49\textwidth}
        \centering
        \includegraphics[width=0.95\columnwidth]{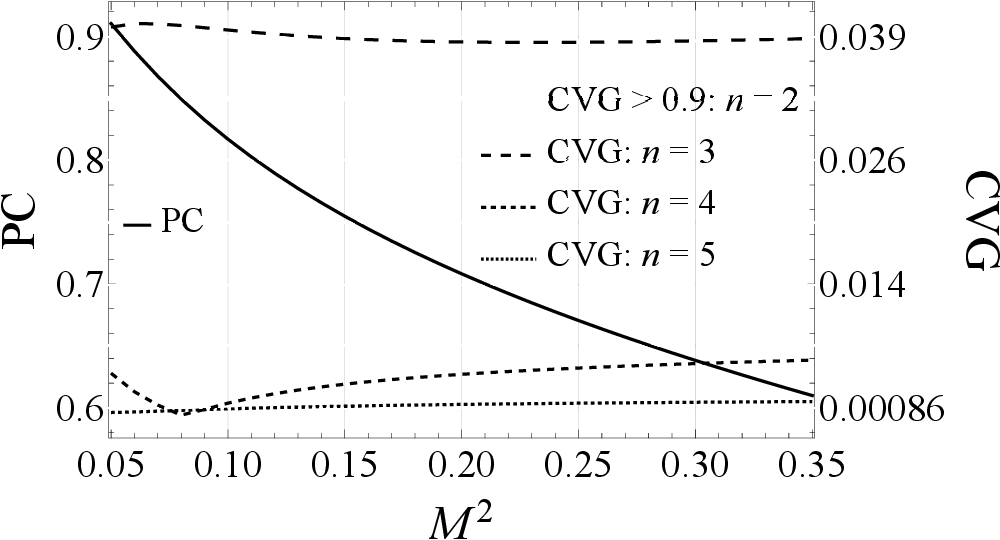}
        \subcaption{P-wave}
    \end{minipage}
    \caption{The pole contribution ratio (PC) and OPE convergence (CVG), shown as 
    a function of the Borel mass $M$. }
    \label{Fig:PCCVG}
\end{figure*}

To extract the resonance pole in Eq.~\eqref{eq:EPget}, the main task is to determine two parameters $\alpha$ and $\theta_P$.
In practice, what we can calculate are $A'+\alpha B'$ and $C'+\alpha D'$ with changing cancellation parameter $\alpha$, scanning angle $\theta_0$, threshold $r_\mathrm{th}$ and Borel mass $M$.
The ranges of $M$ and $r_\mathrm{th}$ are constrained by the conventional QCD sum rules criteria of PC and CGV, as discussed below. 
For $\alpha=-f_x/f_y$ and $\theta_0=\theta_P$, we propose the following method to determine their values.

For the correct $\alpha$ satisfying $f_x+\alpha f_y=0$ and for $\theta_0\approx \theta_P$, the combination $A'+\alpha B'$ in Eq.~\eqref{eq:abprime} becomes independent of $r_{\mathrm{th}}$ with fixed $M^2$.
This can be seen by expanding the integrand in Eq.~\eqref{eq:Tph},
\begin{align}
&(A'+\alpha B')^\mathrm{ph}\nonumber\\
\approx&\int_{0}^{r_{\mathrm{th}}}\dd r\left\{\frac{f_x+\alpha f_y}{r-r_P}+\frac{(\alpha f_x-f_y)r_P(\theta_P-\theta_0)}{(r-r_P)^2+rr_P(\theta_P-\theta_0)^2}\right\}\nonumber\\
&\times W(r\ee^{\ii\theta_0},M)+\mathcal{O}[(\theta_P-\theta_0)^2]+\mathrm{possible\ residue}.\label{eq:abprimeexpand}
\end{align}
For $\alpha=-f_x/f_y$ and $\theta_0\approx\theta_P$, the $1/(r-r_P)$ term cancels, and a $\delta(r-r_P)$ function appears as $\frac{r0^\pm}{(r-r_P)^2+(r0^\pm)^2}=\pm\pi\delta(r-r_P)$, yielding the $r_{\mathrm{th}}$-independence. 
This property allows one to determine $\alpha$ and $\theta_0$.
Furthermore, whether the value of ``possible residue" is zero depends on whether the considered singularity is within the contour, which solely relies on the magnitude relationship between $\theta_0$ and $\theta_P$, and is itself independent of the numerical value of $r_{\mathrm{th}}$.
In addition, the above discussion in principle is based on a fixed Borel mass, thus, we can pick out the specific values of $\alpha$ and $\theta_0$ for each Borel mass.
However, for a properly defined Borel window, the optimal values of $\alpha$ and $\theta_0$ turn out to be rather stable for the square well potential, which will be shown later.

The $r_{\mathrm{th}}$ independence of $A'+\alpha B'$ motivates us to formulate the following function to search for $\alpha$ and $\theta_0$.
\begin{align}
&\sigma_1(\alpha,\theta_0)\equiv
\mathrm{STD}[(A'+\alpha B'),\{r_{\mathrm{th}1},\cdots,r_{\mathrm{th}n}\}],\label{eq:sigma1}
\end{align}
where $r_{\mathrm{th}i}=r_{\mathrm{th}}+i\Delta r_\mathrm{th},\ i\in\mathbb{Z}$ and the statistical standard deviation(STD) is defined as
\begin{align}
&\mathrm{STD}[Y,{\{X_1,\cdots,X_n\}}]=\sqrt{\frac{\sum_i^n[Y_i(X_i)-\bar{Y}]^2}{n-1}},
\end{align}
with $\bar{Y}=(1/n)\sum_{i}^nY_i(X_i)$.
We determine $\alpha$ and $\theta_0$ by minimizing $\sigma_1$.
Furthermore, in Appendix~\ref{appendix:sigma2}, we propose another statistical standard deviation $\sigma_2$, for $B^\prime-\alpha A^\prime$, which can further confirm the robustness of the values of $\alpha$ and $\theta_0$ obtained from $\sigma_1$. 

Notably, we should point out that the intrinsic uncertainties arise from higher order corrections: 
(i) $\mathcal{O}[(\theta_P-\theta_0)^2]$ terms in the expansion of $T^\mathrm{ph}(E)$, 
(ii) $\mathcal{O}[(r_P^\frac{\pi}{\pi-\theta_0}-r_0^\frac{\pi}{\pi-\theta_0})(\theta_P-\theta_0)/M^2)]$ corrections from the Gaussian weight $W(r_P\ee^{\ii\theta_P},M)-W(r_P\ee^{\ii\theta_0},M)$, 
and (iii) contributions from background in $T^\mathrm{ph}(E)$ such as excited states and continuum.  
These yield deviations that affect the precision of the extracted $\alpha$ and related parameters.

After fixing the optimal $\alpha$ and $\theta_0$, we next specify the conditions for the Borel window $[M_\mathrm{min},M_\mathrm{max}]$, 
which is chosen to ensure the pole dominance and the OPE convergence. 
To suppress contributions from excited states and the continuum, we require that the pole contribution, 
\begin{equation}\label{PCcriterion}
    \mathrm{PC}=\frac{\left|(A'+\alpha B')^{\mathrm{OPE}(n)}\right|}{\left|(A'+\alpha B')^{\mathrm{OPE}(n)}\right|+\mathcal{K}},
\end{equation}
should be sufficiently large, where 
\begin{equation}
\mathcal{K}=\left(\int^{r_{\mathrm{th}}\ee^{\ii\pi}}_{-\infty}+\int_{r_{\mathrm{th}}\ee^{\ii\theta_0}}^{\infty\ee^{\ii\theta_0}}\right)|T^\mathrm{OPE}(E)|W(E,M)\mathrm{d} E,
\end{equation} 
represents background contribution from excited states and the continuum.
Meanwhile, the OPE convergence is qualified by 
\begin{equation}\label{CVGcriterion}
   \mathrm{CVG}_n=\left|1-\frac{(A'+\alpha B')^{\mathrm{OPE}(n-1)}}{(A'+\alpha B')^{\mathrm{OPE}(n)}}\right|.
\end{equation}
Since the $A'+\alpha B'$ is almost independent of the threshold $r_{\mathrm{th}}$ for the optimal $\alpha$ and $\theta_0$,
the last variable $r_P$ will also be weakly dependent on the threshold $r_{\mathrm{th}}$. %
Namely, any $r_{\mathrm{th}}$ in the proper range will result in the similar pole position and residue.
Nevertheless, we still need to verify the final results by their dependence on the Borel mass. 
The dependence of $|E_P^\mathrm{ex}|$ on the Borel mass can be quantified by
\begin{equation}\label{XiSquared}
\chi^2(r_{\mathrm{th}})=\int_{M_\mathrm{min}^2}^{M_\mathrm{max}^2}\mathrm{d}M^2\frac{[E_P^\mathrm{ex}(M^2,\mathbb{C}^-_{\mathrm{arc}:r_\mathrm{th}})-\bar{E}_P^\mathrm{ex}]^2}{M_\mathrm{max}^2-M_\mathrm{min}^2},
\end{equation}
where the average energy is
\begin{equation}
\bar{E}_P^\mathrm{ex}=\int_{M_\mathrm{min}^2}^{M_\mathrm{max}^2}\mathrm{d}M^2\frac{E_P^\mathrm{ex}(M^2,\mathbb{C}^-_{\mathrm{arc}:r_\mathrm{th}})}{M_\mathrm{max}^2-M_\mathrm{min}^2},
\end{equation} 
ensuring a stable Borel plateau.

Finally, the parameter $\omega_0=\omega(E_0,\theta_0)$ with $E_0=r_0\ee^{\ii\theta_0}$ (Eq.~\eqref{eq:Etoomega}) can be chosen iteratively: 
an initial guess $E_{0}$ (1st iteration) close to the expected pole is refined by updating $E_{0}$ ($m$-th iteration)$\approx E_P^\mathrm{ex}$ (($m-1$)-th\ iteration) until convergence.
Note that the initial $E_0$ does not need to coincide with the pole position.

\section{Examples Extracting the Complex pole Position}
\label{sec:example}

As a first application of the proposed method, we employ the square-well potential to extract the resonance poles explicitly. 
The results are presented for both $S$- and $P$-wave channels.

\subsection{Determination of the optimal $\alpha$ and $\theta_0$}

We apply the proposed procedure to the $S$-wave case with $z_0=\sqrt{2V_0}=0.4\pi$ and the $P$-wave case with $z_0=\sqrt{2V_0}=0.9\pi$, using the contour shown in Fig.~\ref{Fig:ContourIntegral2}.
At large energies $E$, the $T_0(E)$ ($S$-wave) and $T_1(E)$ ($P$-wave) can be approximated by their truncated expansions $T^{\mathrm{OPE}(n=5)}_l(E)\approx T_l(E)$ along the complex arc up to truncation order $n=5$, as shown in Fig.~\ref{Fig:ScatteringAmplitudeAndWeightFunction}. 
To improve convergence, expansions are performed at $V_0=0.7\lesssim 0.4\pi$ for the $S$-wave and $V_0=3.9 \lesssim 0.9\pi$ for the $P$-wave cases, rather than at $V_0=0$. 

Under the mapping transformation of Eq.~\eqref{eq:Etoomega}, the Gaussian weight function $W(E,M;E_0,\theta_0)$ in Eq.~\eqref{eq:Etowf} is employed in the contour integrals. 
For the $S$-wave, we choose $E_{0}=11.00\,\ee^{\ii\theta_0}$ with $\theta_0=-0.940$, while for the $P$-wave we adopt $E_{0}=0.30\,\ee^{\ii\theta_0}$ with $\theta_0=-0.500$. 
As shown in Fig.~\ref{Fig:ScatteringAmplitudeAndWeightFunction}, the weight functions effectively enhance the pole contributions while suppressing the non-pole background.

To evaluate $(\sigma_1)_{\min}$, the parameter space of $r_\mathrm{th}$ is discretized into $21$ equal steps for two different choices of the $r_\mathrm{th}$ range. 
We employ the ranges $[11.00, 12.00]$ and $[11.00, 13.00]$ for the $S$-wave, and $[0.40,0.50]$ and $[0.40, 0.60]$ for the $P$-wave.
Fig.~\ref{Fig:ThetaAndSigma} shows the minimal value of $\sigma_1$ by varying $\alpha$ at each fixed $\theta_0$. 
The different curves correspond to different choices of the $r_\mathrm{th}$ range and the Borel mass $M$. 
Then, from the minima of these curves, 
we obtain the optimal $\theta_0$ value.
Fig.~\ref{Fig:ThetaAndSigma} shows that while the curves vary, the location of the minima are quite consistent, demonstrating that the extracted $\theta_0$ for $S$- and $P$-wave can be rather stable for different choices of the $r_{\mathrm{th}}$ range.
The results with different $M$ are also consistent with each other, showing that the minimum depends little on the Borel mass. 
In Fig.~\ref{Fig:alphaandsigma}, we show the behavior of the minimum $\sigma_1$ value as a function of $\alpha$ at the optimal $\theta_0$.
This valley structure of $\sigma_1$ identifies the optical $\alpha$.
With this approach, we are able to fix the best values of $\alpha\approx -f_x/f_y$ and $\theta_0\approx\theta_P$, which are independent of the choices of $r_{\mathrm{th}}$ and $M^2$.

\subsection{Borel window and plateau}

In this subsection, we determine the Borel window by the PC and CVG criteria. 
We choose $M^2\in[0.80,2.00]$ with $90\%\gtrsim \mathrm{PC}\gtrsim 80\%$ for the $S$-wave, and $M^2\in[0.10,0.30]$ with $80\%\gtrsim \mathrm{PC}\gtrsim 60\%$ for the $P$-wave, which satisfy the large pole contribution and good convergence. 
Fig.~\ref{Fig:PCCVG} shows that the OPE convergence is ensured for $n\approx5$.

\begin{figure*}
    \centering
    \begin{minipage}{0.49\textwidth}
        \centering
        \includegraphics[width=0.8\columnwidth]{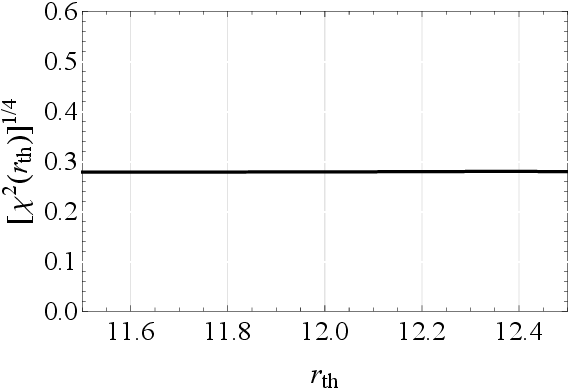}
        \subcaption{S-wave}
    \end{minipage}
    \begin{minipage}{0.49\textwidth}
        \centering
        \includegraphics[width=0.8\columnwidth]{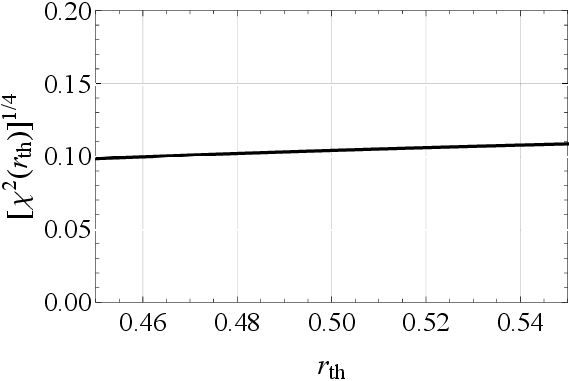}
        \subcaption{P-wave}
    \end{minipage}
    \caption{Dependence of $\chi^2(r_\mathrm{th})$ on the threshold parameter $r_\mathrm{th}$.}
    \label{Fig:rth}
\end{figure*}

Since when determining the minimum $\sigma_1$, we require the least dependence of $A'+\alpha B'$ on the threshold $r_{\mathrm{th}}$, $(r_P^\mathrm{ex})^{\frac{\pi-\theta_0}{\pi}}=(C'+\alpha D')/(A'+\alpha B')$ and even the final $\chi^2(r_\mathrm{th})$ are hardly dependent on the threshold, as demonstrated in Fig.~\ref{Fig:rth} to confirm the stable Borel plateau. 
Here we choose the median value $r_\mathrm{th}=12.00$ for $S$-wave and  $r_\mathrm{th}=0.50$ for $P$-wave as the optimal thresholds within the range of their threshold sets.

\begin{figure*}
    \centering
    \begin{minipage}{0.49\textwidth}
        \centering
        \includegraphics[width=0.8\columnwidth]{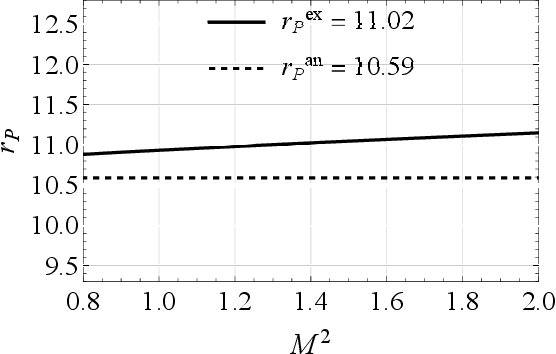}
        \subcaption{S-wave}
    \end{minipage}
    \begin{minipage}{0.49\textwidth}
        \centering
        \includegraphics[width=0.8\columnwidth]{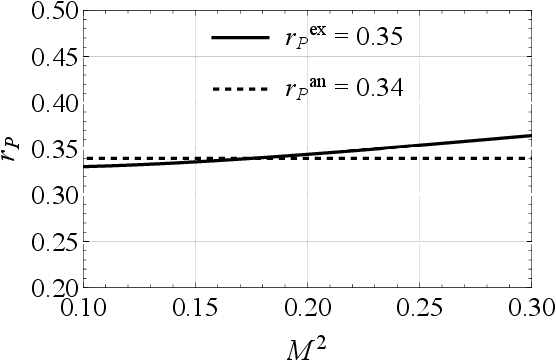}
        \subcaption{P-wave}
    \end{minipage}
    \caption{The pole radius $r_P$, shown as a function of the (squared) Borel mass $M^2$. }
    \label{Fig:rPExtraction}
\end{figure*}

Finally, the pole positions $E_P^\mathrm{ex}$ and the residues $F^\mathrm{ex}$ are extracted via Eq.~\eqref{eq:EPget} and Eq.~\eqref{eq:FExtraction} for both $S$- and $P$-wave. 
The resulting Borel plateau shown in Fig.~\ref{Fig:rPExtraction}, demonstrate that $(r_P^\mathrm{ex})^{S-\mathrm{wave}}\approx11.02$ and $(r_P^\mathrm{ex})^{P-\mathrm{wave}}\approx0.35$ are nearly independent of $M^2$. 
The obtained parameters and their comparison with the analytic results are summarized in Table~\ref{tab:ParametersComparison}.

\begin{table}[h!]
    \newcommand{\arrayschretch}{3}
    \centering
    \caption{Comparison between the analytical results and extracted complex pole positions for the $S$-wave and  $P$-wave resonances. $n=5$ is chosen for OPE.}
    \setlength{\tabcolsep}{10pt}
    \renewcommand{\arraystretch}{2}
    \begin{tabular}{ccc}\bottomrule[2pt]
       Parameters   & S-wave & P-wave \\
         \bottomrule[1pt]
        Threshold $r_{\mathrm{th}}$ & $12.00$ & $0.50$ \\
        Optimal $\theta_0$ & $-0.948$ & $-0.560$ \\
        Optimal $\alpha$ &  $-2.38$ & $0.14$ \\
        $r_0$ & $11.00$ & $0.30$ \\
        Borel window $M^2$ & $0.70\sim2.00$ & $0.10\sim0.30$ \\
        Pole Contribution & $90\%\sim80\%$ & $80\%\sim60\%$ \\
         \bottomrule[1pt]
        Pole extracted $E_P^\mathrm{ex}$ & $11.02\,\ee^{-\ii\,0.948}$ & $0.35\,\ee^{-\ii\,0.560}$ \\
        Pole analytic $E_P^\mathrm{an}$ & $10.59\,\ee^{-\ii\, 0.936}$ & $0.34\,\ee^{-\ii\, 0.552}$ \\ 
        Residue extracted & $0.19+0.08\,\ii$ & $-0.02+0.17\,\ii$ \\
        Residue analytic & $0.24+0.03\,\ii$ & $ -0.04+0.18\,\ii$ \\ 
        \bottomrule[2pt]
    \end{tabular}
    \label{tab:ParametersComparison}
\end{table}

\section{Summary and Discussion}
\label{sec:concl}

In this paper, we have developed a novel method, the ``resonance sum rule'' using Cauchy’s theorem to extract resonance poles on the second Riemann sheet of the correlation function within the QCD sum rule framework. 
Although the core idea is conceptually straightforward, its application requires overcoming several technical challenges. 

To demonstrate the validity and effectiveness of our approach, we first applied it to a toy model: the quantum mechanical scattering problem with a square-well potential, which can be solved analytically. 
Focusing on the first resonance poles of the $S$-wave and $P$-wave scattering amplitudes in an attractive square-well potential, we demonstrated that the pole position and the residue can be reliably extracted from the resonance sum rule. 

To achieve this, we employ a contour integral of the scattering amplitude weighted by a ``Borel function". 
The contour is constructed to simultaneously enclose the resonance pole on the second Riemann sheet and the high-energy region where the $1/z$ expansion (with $z=\sqrt{2E}$), mimicking the OPE, is valid. 
By rotating the contour around $E=0$ onto the second Riemann sheet and varying the rotation angle and the contour radius, one can extract the resonance pole.

It is found that this method works effectively in the square-well potential model. 
For both $S$- and $P$-wave channels, the pole positions and residues can be extracted from the resonance sum rule. 
In particular, we utilize the sensitivity of the sum rule to the rotation angle in order to determine the ratio of the real and imaginary parts of the pole position. 
The results are consistent with the analytic results.
For the pole positions and residues, the error is within $5\%$ and $20\%$, respectively.
The precision is ultimately limited by the validity of the $1/z$ expansion, which reflects the analogous limitations of the OPE in QCD sum rules.

To extend this method to real hadron resonances, further investigations and refinements are required. 
As a next step, we plan to test the method on the $\rho$ meson resonance. 
In principle, it should be possible to carry out such an analysis to extract the complex pole from the $P$-wave $\pi\pi$ scattering amplitude on the second Riemann sheet.


\begin{acknowledgements}

This work is supported in part by JSPS Kakenhi No.JP20K03959 (MO, PG), JP23K03427 (MO, WGJ), JP25H00400 (PG), JP22H00122 (PG) and JP24K17055 (WGJ), and by the National Natural Science Foundation of China (NSFC) under Grants Nos. 12175239 and 12221005 (JJW), and by the Chinese Academy of Sciences under Grant No. YSBR-101(JJW).
\end{acknowledgements}

\appendix

\section{Derivation of various integrals}

\label{Appendix:Phenomenological Derivation}

The definition of the phenomenological scattering amplitude with $E=r\ee^{\ii\theta}$ is
\begin{equation}
    T^\mathrm{ph}(E)=\frac{F}{E-E_P}+\mathrm{background},
\end{equation}
where $E=r\ee^{\ii\theta},\ \ E_P=r_P\ee^{\ii\theta_P}$ and $F=f_x+\ii f_y$. If $\theta\approx\theta_P$ one has
\begin{align}
&T^\mathrm{ph}(E)\nonumber\\
&=\frac{F\ee^{-\ii\theta}}{r-r_P\ee^{\ii(\theta_P-\theta)}}
=\frac{F\ee^{-\ii\theta}(r-r_P\cos{\delta}+\ii r_P\sin{\delta})}{(r-r_P\cos{\delta})^2+(r_P\sin{\delta})^2}\nonumber,\\
&=F\ee^{-\ii\theta}\left[\frac{1}{r-r_P}+\frac{\ii r_P\delta}{(r-r_P)^2+rr_P\delta^2}\right]+\mathcal{O}(\delta^2)\nonumber,\\
&\xlongequal[]{\delta\to0} 
F\ee^{-\ii\theta}\left[\frac{1}{r-r_P}\pm\ii\pi\delta(r-r_P)\right],
\end{align}
where $\delta\equiv\theta_P-\theta$ and $+(-)$ relates to $\delta>0$ ( $\delta<0$ ). 

Eq.~(\ref{Equ:APlusAlphaB}) is then derived as
\begin{equation}\begin{split}
    &(A+\alpha B)^\mathrm{ph}
    \\
    &\equiv \left(\int^0_{r_{\mathrm{th}}\ee^{\ii\pi}}+\int_0^{r_{\mathrm{th}}\ee^{\ii\theta_0}}\right)T^\mathrm{ph}(E)W(E,M)\dd E\\
    &\xlongequal[]{\mathrm{suppress\ }\int_{r_{\mathrm{th}}\ee^{\ii\pi}}^{0}}\int_{0}^{r_{\mathrm{th}}}\Bigg[\frac{f_x+\alpha f_y}{r-r_P}\\
    &\quad+\frac{(\alpha f_x-f_y)r_P(\theta_P-\theta_0)}{(r-r_P)^2+rr_P(\theta_P-\theta_0)^2}\Bigg]W(E,M)\mathrm{d}r\\
    &\xlongequal[]{\theta_0\to\theta_P}\int_{0}^{r_{\mathrm{th}}}\Bigg[\frac{f_x+\alpha f_y}{r-r_P}\\
    &\quad\pm\pi(\alpha f_x-f_y)\delta(r-r_P)\Bigg]W(r\ee^{\ii\theta_0},M)\mathrm{d}r\\
    &=\left\{\begin{array}{cc}
-\pi(\alpha f_x-f_y)W(r_P\ee^{\ii\theta_0},M) &
\,\,\,\, \theta_P<\theta_0\\
\pi(\alpha f_x-f_y)W(r_P\ee^{\ii\theta_0},M) & \,\,\,\,  \theta_0<\theta_P
\end{array}\right.,\nonumber
\end{split}
\end{equation}
where the superscript ``$\mathrm{ph}$" denotes that $T^{\mathrm{ph}}(E)$ is used to approximate the integral. 
Similarly, the linear combination $C+\alpha D$ in Eq.~(\ref{Equ:CPlusAlphaD}) is
\begin{equation}\begin{split}
&(C+\alpha D)^\mathrm{ph}_{\theta_0\to\theta_P}\xlongequal[]{\mathrm{suppress\ }\int_{r_{\mathrm{th}}\ee^{\ii\pi}}^{0}}\int_{0}^{r_{\mathrm{th}}}\Bigg[\frac{f_x+\alpha f_y}{r-r_P}\\
&\quad\pm\pi(\alpha f_x-f_y)\delta(r-r_P)\Bigg]W(r\ee^{\ii\theta_0},M)r^\frac{\pi}{\pi-\theta_0}\mathrm{d}r\\
&=\left\{\begin{array}{cc}
-\pi(\alpha f_x-f_y)W(r_P\ee^{\ii\theta_0},M)r_P^\frac{\pi}{\pi-\theta_0} &\,\,\,\,\theta_P<\theta_0\\
\pi(\alpha f_x-f_y)W(r_P\ee^{\ii\theta_0},M)r_P^\frac{\pi}{\pi-\theta_0} &\,\,\,\,\theta_0<\theta_P
\end{array}\right..\nonumber
\end{split}
\end{equation}
Note that in both cases we enforce the condition $f_x+\alpha f_y = 0$ to eliminate the $1/(r-r_P)$ term.

With the residue of the closed contour integral in Eq. \eqref{eq:TCinte}, 
\begin{align}\begin{split}
    &\oint_{\mathbb{C}}T^\mathrm{ph}(E)W(E,M)\mathrm{d}E
    \nonumber\\
    &=\left\{\begin{array}{cc}
        0 & \,\,\,\,\theta_P<\theta_0  \\
        2\pi\ii FW(r_P\ee^{\ii\theta_P},M) & \,\,\,\,\theta_0<\theta_P
    \end{array}\right.,
\end{split}\end{align}
the arc integral reads
\begin{equation}\label{appendixA:A'+B'}\begin{split}
    &(A'+\ii B')^\mathrm{ph}_{\theta_0\to\theta_P}=\int_{\mathbb{C}_\mathrm{arc}^-}T^\mathrm{ph}(E)W(E,M)\mathrm{d}E\\
    &=(A+\ii B)^\mathrm{ph}_{\theta_0\to\theta_P}-\left\{\begin{array}{cc}
        0 &\,\,\,\, \theta_P<\theta_0 \\
        2\pi\mathrm{i}FW(r_P\ee^{\ii\theta_P},M)&\,\,\,\, \theta_0<\theta_P
    \end{array}\right.,
\end{split}\end{equation} 
where the explicit forms of $A'$ and $B'$ are
\begin{equation}
A'^{\,\mathrm{ph}}_{\theta_0\to\theta_P}=A^\mathrm{ph}_{\theta_0\to\theta_P}+\left\{\begin{array}{cc}
    0&\,\,\,\,\theta_P<\theta_0 \\2\pi f_yW(r_P\ee^{\ii\theta_P},M)& \,\,\,\,\theta_0<\theta_P\end{array}\right.,
\end{equation}
and
\begin{equation}
B'^{\,\mathrm{ph}}_{\theta_0\to\theta_P}
=B^\mathrm{ph}_{\theta_0\to\theta_P}+\left\{\begin{array}{cc}
    0& \,\,\,\,\theta_P<\theta_0\\-2\pi f_xW(r_P\ee^{\ii\theta_P},M)& \,\,\,\,\theta_0<\theta_P\end{array}\right..
\end{equation}
This yields the simple linear combination,
\begin{equation}\begin{split}
    (A'+\alpha B')^\mathrm{ph}_{\theta_0\to\theta_P}=-\pi(\alpha f_x-f_y)W(r_P\ee^{\ii\theta_P},M).
\end{split}\end{equation}
%
$\alpha$ is still chosen as $-f_x/f_y$ to cancel the contribution from the $1/(r-r_P)$ term. 
Similarly, the definitions of $C'$ and $D'$ 
are given as
\begin{align}
    &(C'+\ii D')^\mathrm{ph}_{\theta_0\to\theta_P}=\int_{\mathbb{C}_\mathrm{arc}^-}T^\mathrm{ph}(E)W(E,M)\omega(E,\theta_0)\mathrm{d}E\nonumber\\
    &=(C+\ii D)^\mathrm{ph}_{\theta_0\to\theta_P}\nonumber\\
    &\quad-\left\{\begin{array}{cc}
        0 &, \theta_P<\theta_0 \\
        2\pi\ii FW(r_P\ee^{\ii\theta_P},M)r_P^\frac{\pi}{\pi-\theta_0}&, \theta_0<\theta_P
    \end{array}\right.,\label{appendixA:C'+D'}
\end{align}
which ultimately results in the linear combination
\begin{equation}\begin{split}
     ( C'+\alpha D')^\mathrm{ph}_{\theta_0\to\theta_P}=-\pi(\alpha f_x-f_y)W(r_P\ee^{\ii\theta_P},M)r_P^\frac{\pi}{\pi-\theta_P}.
\end{split}\end{equation}

\section{Additional variable for extracting $\alpha$}
\label{appendix:sigma2}

Here we propose a new variable $B'-\alpha A'$ that can also help to extract the correct $\alpha=-f_x/f_y$.
In the $M\rightarrow \infty$ limit, when $\theta_0\sim \theta_P$, $B'-\alpha A'$ can be expressed as
\begin{align}
&\left.(B'-\alpha A')^\mathrm{ph}\right|_{M^2\to\infty}\nonumber\\
\approx&\int_{0}^{r_{\mathrm{th}}}\dd r\left\{\frac{f_y-\alpha f_x}{r-r_P}+\frac{(f_x+\alpha f_y)r_P(\theta_P-\theta_0)}{(r-r_P)^2+rr_P(\theta_P-\theta_0)^2}\right\}\nonumber\\
&\times W(r\ee^{\ii\theta_0},M)+\mathcal{O}[(\theta_P-\theta_0)^2]+\mathrm{possible\ residue}.\label{eq:abprimeexpand}
\end{align}

Then, if $\alpha=-f_x/f_y$, we find that $\left.(B'-\alpha A')^\mathrm{ph}\right|_{M^2\to\infty}$ becomes independent of $\theta_0$ for any fixed $r_{\mathrm{th}}$.
Thus, we can define a new statistical standard deviation as 
\begin{align}
&\sigma_2(\alpha,r_\mathrm{th})\equiv
\mathrm{STD}[(B'- \alpha A'),\{\theta_{01},\cdots,\theta_{0n}\}],\label{eq:sigma2}
\end{align}
with $\theta_{0i}=\theta_0+i\Delta\theta_0$. 

\begin{figure}[h!]
    \centering
    \includegraphics[width=0.9\linewidth]{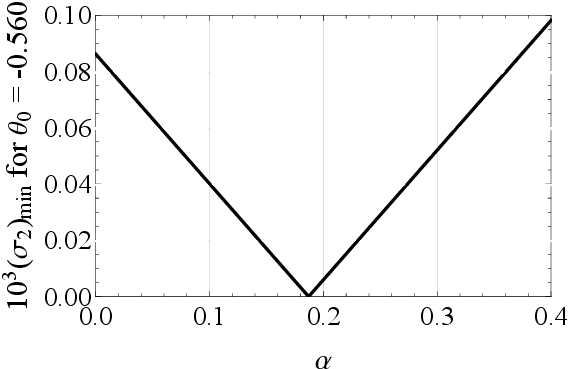}
    \caption{The value of $\sigma_2$ as a function of $\alpha$ with the fixed $r_\mathrm{th}$ and $\theta_0$.}
    \label{Fig:alphaandsigma_APPENDIX}
\end{figure}
As a consequence, $\sigma_2$ enables us to determine $\alpha$ by minimizing $\sigma_2$ with fixed $r_{\mathrm{th}}$ and $\theta_0$.
%
To exemplify, for the $P$-wave case we take $r_{\mathrm{th}}=0.50$, $M^2=10$, and $\theta_0=-0.560+i\Delta\theta_0$ where $i\in\{j|j\in\mathbb{Z}\mathrm{\ and\ }j\in[-10,10]\}$ and $\Delta\theta_0=10^{-4}$. $\sigma_2$, as a function of $\alpha$ is given in Fig.~\ref{Fig:alphaandsigma_APPENDIX}, indicating a minimum for the optimal $\alpha^\mathrm{extracted}=0.19$ that reconfirms the existence of a resonance pole with a residue 
consistent with the method presented in the main text.

\bibliography{biblio.bib}

\end{document}